\documentclass[aps,twocolumn,showpacs,superscriptaddress,groupedaddress,amsmath,amssymb]{revtex4}  
\usepackage{aas_macros} 
\usepackage{graphicx}   
\usepackage{dcolumn}    
\usepackage{bm}         
\usepackage{makecell}   
\usepackage{mathrsfs}   
\usepackage{color}      
\usepackage[version=4]{mhchem}

\newcommand{\bdf}{$\beta$df}                               
\newcommand{\mcbdf}{mc-$\beta$df}                          

\graphicspath{{./}{figs/}}

\begin{document}

\hspace{5.2in} \mbox{LA-UR-21-31648}

\title{$\beta^{-}$-delayed fission in the coupled Quasi-particle Random Phase Approximation plus Hauser-Feshbach approach}

\author{M.~R.~Mumpower}
\email{mumpower@lanl.gov}
\homepage[]{https://www.matthewmumpower.com}
\affiliation{Theoretical Division, Los Alamos National Laboratory, Los Alamos, NM 87545, USA}

\author{T.~Kawano}
\affiliation{Theoretical Division, Los Alamos National Laboratory, Los Alamos, NM 87545, USA}

\author{T.~M.~Sprouse}
\affiliation{Theoretical Division, Los Alamos National Laboratory, Los Alamos, NM 87545, USA}

\date{\today}

\begin{abstract}
Beta-delayed neutron emission and $\beta$-delayed fission (\bdf) probabilities were calculated for heavy, neutron-rich nuclei using the Los Alamos coupled Quasi-Particle Random Phase Approximation plus Hauser-Feshbach (QRPA+HF) approach. 
In this model, the compound nucleus is initially populated by $\beta$-decay and is followed through subsequent statistical decays taking into account competition between neutrons, $\gamma$-rays and fission. 
The primary output of these calculations includes branching ratios along with neutron and $\gamma$-ray spectra. 
We find a relatively large region of heavy nuclides where the probability of \bdf\ is near 100\%. 
For a subset of nuclei near the neutron dripline, delayed neutron emission and the probability to fission are both large which leads to the possibility of multi-chance \bdf\ (\mcbdf). 
We comment on prospective neutron-rich nuclei that could be probed by future experimental campaigns and provide a full table of branching ratios in ASCII format in the supplemental material for use in various applications. 
\end{abstract}

\maketitle

\section{Introduction}
Beta-delayed fission (\bdf) is a two step nuclear decay process in which fission follows electron capture (EC), $\beta^{+}$ decay or $\beta^{-}$ decay. 
Discovered by experiments in the 1960's, this process starts with a parent nucleus which undergoes EC or $\beta^{+/-}$ decay to an excited state in the daughter nucleus. 
The populated state may fission if the excitation energy is near or greater than the fission barrier of the daughter nucleus and is in competition with particle emission and $\gamma$ de-excitation. 
This rare decay mode in near-stable isotopes is limited by the relatively small $\beta$-decay Q-value of the parent nucleus making it a unique probe of low energy structure of atomic nuclei. 
In nuclei with extreme neutron-excess this decay mode may be more prevalent, playing a role in the cosmos by influencing the formation of the elements found on the periodic table in astrophysical events \cite{Kodama+75}. 

Difficulty in the production of exotic nuclei which may undergo fission coupled with small branching ratios relative to other processes (such as $\alpha$-decay) makes the investigation of \bdf\ very challenging in a laboratory environment \cite{Andreyev+13}. 
Just under thirty cases of \bdf\ on both the neutron-deficient and neutron-rich side of stability have been studied experimentally with most of this progress coming in recent years at radioactive beam facilities. 
The bulk of current data resides on the neutron-deficient side of stability, e.g. most recently the two isotopes of thallium ($Z=81$), $^{178}$Tl and $^{180}$Tl have been shown to be the lighest measured \bdf\ precursors \cite{Andreyev+10, Liberati+13, Elseviers+13}. 
While the \bdf\ branching ratio of the neutron-deficient isotope $^{242}$Es at $0.6$(2) $\times$ $10^{-2}$ remains one of the largest measured to date, see Table 1 of Ref. \cite{Andreyev+13} for full list. 
The partial \bdf\ half-lives of neutron-deficient nuclei have been found to follow a systematic trend \cite{Ghys+15}, however further measurements are required to determine if the systematics can be extended to neutron-rich nuclei. 
Measurements have been made on only six neutron-rich nuclei, all of which exhibit extremely small branchings relative to measurements of neutron-deficient nuclei. 
The reason for this observation comes from the fact that experiments cannot yet reach nuclei where the fission barrier height ($B_{\textrm f}$) is sufficiently smaller than $Q_{\beta^{-}}$. 
Such low branching ratios are beyond the fidelity of any current nuclear model prediction. 

Excluding the paucity of experimental data, the description of \bdf\ persists as an open challenge to theory. 
This stems from the requirement of an all encompassing description of the complexity of the atomic nucleus. 
To list a few examples: the $\beta$-strength function, excited states, single particle and collective effects as well as a fission properties that depend on the nuclear potential-energy landscape, and traversal through it (resulting in the production of fission fragments) must all be modeled. 
Of particular interest to the description of \bdf\ are those neutron-rich nuclei that may participate in the astrophysical rapid neutron capture or $r$-process of nucleosynthesis \cite{Mumpower+16r}. 
While these nuclei remain out of reach to experimental facilities, heavy $r$-process nuclei have relatively large $Q_{\beta^{-}}$ compared to $B_{\textrm{f}}$ which is a favorable condition for large \bdf{} branching ratios. 
The work of Thielemann \textit{et al.} \cite{Thielemann+83} laid the groundwork for the theoretical description of \bdf\ and its application to the rapid neutron capture or $r$ process of nucleosynthesis. 
Statistical approaches, as studied here, offer further exploration of delayed fission phenomenon \cite{Mumpower+18a, Minato+21}. 

In this work, we study the \bdf\ of neutron-rich nuclei using the recently developed Quasi-particle Random Phase Approximation plus Hauser-Feshbach (QRPA+HF) framework. 
This microscopic approach starts with a compound nucleus initially populated by $\beta$-decay and follows the subsequent statistical decay taking into account competition between neutrons, $\gamma$-rays and fission. 
We find a relatively large region of the chart of nuclides where the probability of \bdf{} is near 100\% that prevents the production of superheavy elements in nature \cite{Mumpower+18a}. 
The decay chains of very neutron-rich nuclei near the neutron dripline exhibit large neutron multiplicity and large probability to fission leading to the possibility of multi-chance \bdf{} or \mcbdf{} for short. 
This decay mode results in multiple fission chances after $\beta$-decay analogous to multi-chance neutron induced fission. 
We discuss the theoretical basis of our model and provide tabulated values in the supplemental material. 

\section{Model}
\begin{figure*}
 \begin{center}
  \centerline{\includegraphics[width=150mm]{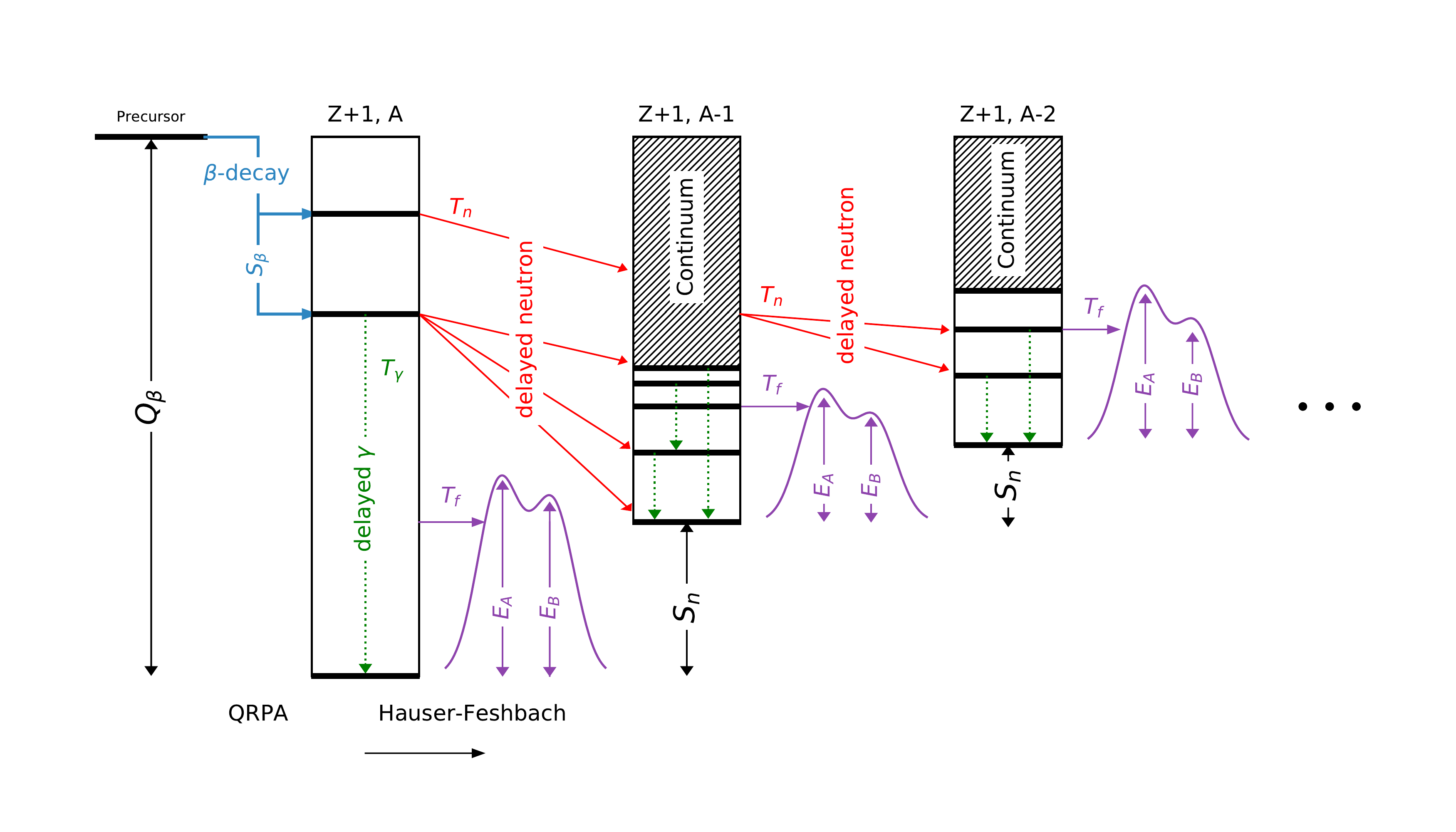}}
  \caption{\label{fig:model} (Color online) Schematic of the combined QRPA+HF approach when applied to \bdf. Initial population of the daughter nucleus ($Z+1$,$A$) is determined by the $\beta$-strength function ($S_\beta$) using QRPA. The competition between neutron emission, $\gamma$ de-excitation and fission are then handled in the HF framework for which the transmission coefficients ($T_{\textrm n}$, $T_\gamma$, $T_{\textrm f}$) are calculated respectively at each stage of the statistical decay. In some cases, the fission barrier heights ($E_A$, $E_B$) may contain only a single hump ($E_B=0$), and could possibly be much larger than denoted by this schematic. Multi-chance \bdf\ occurs for the daughter nucleus ($Z+1$,$A-1$) and beyond. The trailing dots denotes the statistical decay continues until the total available excitation energy ($Q_\beta$) is exhausted.}
 \end{center}
\end{figure*}

The Quasi-particle Random Phase Approximation plus Hauser-Feshbach (QRPA+HF) approach was introduced in Refs.~\cite{Kawano+08, Mumpower+16a} using version 3.3.3 of the Los Alamos Hauser-Feshbach statisitical decay code. 
Here, we apply this framework to the calculation of delayed-neutron emission in the presence of $\beta$-delayed fission and now use version 3.5.0. 
Among other modifications, the change in version number represents the modifications to the code to include the description of fission during the statistical decay. 
The calculation of \bdf{} in this approach is a two step process starting with the $\beta$-decay of the precursor nucleus ($Z$,$A$) where $Z$ represents the atomic number, and $A$ is the total number of nucleons, followed by the statistical decay of the subsequent daughter generations, ($Z+1$,$A-j$) where $j$ is the number of neutrons emitted, ranging from $0$ to $10$. 
Ground state properties are taken from the 2012 version of the Finite Range Droplet Model \cite{FRDM2012} unless otherwise noted. 
The procedure for the QRPA+HF calculation is shown schematically in Fig \ref{fig:model} and outlined in detail below. 

The $\beta$-decay of the precursor nucleus provides the initial population of the daughter nucleus ($Z+1$,$A$), and is given by the QRPA formalism of Refs.~\cite{Krumlinde+84, Moller+90, Moller+97, Moller+03}. 
QRPA allows for the calculation of excited state properties by using a small perturbing force. 
We use the latest QRPA calculations from Ref.~\cite{Moller+18a} which provides both the Gamow-Teller (GT) and First-Forbidden (FF) contributions, a notable upgrade in this work compared to our existing global model calculations \cite{Mumpower+16a, Mumpower+18a}. 
In this framework, based on a folded-Yukawa interaction, the $\beta$-decay half-life is calculated via
\begin{equation}
   \frac{1}{T_{1/2}} = \sum_{0\leq E_x \leq Q_\beta} S_\beta(E_x) f(Z,Q_{\beta}-E_x) \ ,
   \label{eqn:Thalf}
\end{equation}
where $S_\beta(E_x)$ is the $\beta$-strength function at excitation energy $E_x$ in the daughter nucleus and $f$ is the Fermi function which takes into account the phase space contribution. 
The energy factor can be approximated as $f\sim (Q_\beta-E_x)^{5}$ so that the low-energy excitations or near ground-state transitions dominate the calculation of the half-life. 
Thus, the output of the QRPA (the $\beta$-strength function) at low energies is of critical importance to understanding this phenomenon \cite{Mustonen+16}. 
Where data exists, we may supplement the theoretical calculation with the experimental data as in Ref.~\cite{Mumpower+16a}, however, we note that in the case of neutron-rich nuclei which undergo \bdf\ there is no data with which to compare or include. 

The QRPA solutions for $\beta$-decay strongly depend on the level structure in the participating nuclei. 
Further, these solutions are strongly peaked depending on the unknown level structure for the nuclei of interest. 
We therefore apply a smoothing procedure to the $\beta$-strength distribution by a Gaussian,
\begin{equation}
  \omega(E_x) = C \sum_k b^{(k)}
     {1 \over{\sqrt{2\pi} \Gamma}}
     \exp\left\{
            - {{ [E^{(k)} - E_x]^2 }\over{2\Gamma^2}}
         \right\},
  \label{eq:omega}
\end{equation}
where $b^{(k)}$ are the branching ratios from the parent state to the daughter states $E^{(k)}$, $E_x$ is the excitation energy of the daughter nucleus, $C$ is a normalization constant, and the Gaussian width is taken to scale proportional to $A^{-0.57}$ for excitations above 2 MeV, as used in Ref.~\cite{Moller+97, Moller+03}. 
The choice in this work results in a spreading of roughly $200$ keV for $A=200$ nuclei, and is roughly equal to the error in the mass model. 
Other Gaussian widths have been used in our previous works, for instance a constant value of $\Gamma=100$ keV in Refs.~\cite{Kawano+08, Mumpower+16a}. 
The choice of the Gaussian width represents an intrinsic uncertainty associated with modern QRPA methods. 

With the population of the first daughter nucleus defined by the $\beta$-strength function from the QRPA calculation, the statistical decay can then be followed where the competition between neutron emission, $\gamma$-rays and fission occurs. 
In this second stage of the calculation, the daughter nucleus is assumed to be in a compound state, which means the nucleus is governed by its overall properties rather than the details of the formation process \cite{Weisskopf+56}. 
This assumption leads to the independent factorization of exit channel probabilities; for our purposes it is a good approximation. 
Independent factorization of exit channel probabilities may be modified in certain situations by taking into account the memory of the entrance channel via correction factors \cite{Bertsch+17}. 

The excited state transitions show in Fig.~\ref{fig:model} can be discrete or in the continuum. 
Since there is no known level data for the extremely neutron-rich nuclei studied here, we rely on the Gilbert-Cameron level density \cite{Gilbert+65}. 
This level density formula connects a Fermi gas model to a constant temperature model at a matching energy with parameters taken from systematics at stability \cite{Kawano+06}. 
Shell corrections are applied to the level density using the common Ignatyuk \textit{et al.} prescription \cite{Ignatyuk+75}. 

To easily reference the compound nucleus in the context of $\beta$-delayed calculations, we define the compound state as $c^{(j)}_k$ where $c^{(j)}$ represents the $j$-th compound nucleus after $j$ neutron emissions and $k$ the index of the excited state in the same nucleus. 
Using this shorthand, $c^{(0)}$ is the daughter nucleus ($Z+1$,$A$), $c^{(1)}$ is the granddaughter nucleus ($Z+1$,$A-1$) and so on. 

The transmission coefficient for $\gamma$ de-excitation in the compound nucleus, $c^{(j)}$, is denoted by vertical dashed lines in Fig.~\ref{fig:model}. 
This quantity is calculated using the definition,
\begin{equation}
 T^{(j)}_{\gamma}(E_\gamma) = 2\pi E_\gamma^{(2L + 1)}f^{(j)}_{XL}(E_\gamma) \ ,
\end{equation}
where the transition occurs with $\gamma$-ray energy $E_\gamma=E_i-E_k$, and $f^{(j)}_{XL}$ is the $\gamma$-ray strength function of multipole type $XL$ for compound nucleus $c^{(j)}$. 
The $\gamma$-ray transmission coefficient is often written $T_{XL}$ as shorthand for the multipolarities (E1, M2, E2, M2, etc.). 
In this work, we use the generalized Lorentzian for the E1 $\gamma$-strength function ($\gamma$SF) \cite{Kopecky+87}. 
Additional low energy enhancements, such as the M1 scissors mode \cite{Larsen+10, Mumpower+17} are not considered here. 

The transmission coefficient for neutron emission between two compound states is represented by diagonal solid lines connecting adjacent compound nuclei in Fig.~\ref{fig:model}. 
The calculation of this quantity occurs between two excited states $E_i$ and $E_{k^\prime}$ in neighboring nuclei and can be written down as
\begin{multline}
 T^{(j+1)}_n(E_i,E_{k^\prime}) = \sum_s\sum_l T_{n_{l_s}}(E_i-S^{(j)}_n-E_{k^\prime}) \ ,
 \label{eqn:Tn}
\end{multline}
where $S^{(j)}_n$ is the one neutron separation energy of $c^{(j)}$ and the summation is over all possible partial waves. 
The difference in energy $E_n=E_i-E_{k^\prime}$ is the energy of the delayed neutron. 
The $T_{n_{l_s}}$ values are given using the Koning-Delaroche global optical potential that is optimized for neutron-rich nuclei \cite{Koning+03, Koning+05, Goriely+08}. 
The prime on the second ($k$) index is a reminder that the energy level is in a different compound nucleus. 

The transmission coefficient for fission is approximated using the Hill-Wheeler formula assuming transmission through a parabolic barrier \cite{Hill+53},
\begin{equation}
 T^{(j)}_{\textrm f}(E) = \frac{1.0}{1.0+\exp(2\pi \frac{B_{\textrm f}-E}{C})} \ ,
 \label{eqn:Tf}
\end{equation}
where $B_{\textrm{f}}$ is the fission barrier height, $C$ is the curvature and $E$ is the relative excitation energy of the $j$-th compound nucleus. 
The fission transmission coefficient is represented as a horizontal solid line to the right of each compound nucleus in Fig.~\ref{fig:model}. 
Fission curvatures are defined separately for even-even, odd-$A$ and odd-odd nuclei as in Ref.~\cite{Thielemann+83}. 
The effective transmission coefficient for fission depends on the number of fission barriers. 
For instance, with two barriers,
\begin{equation}
 T^{eff}_{\textrm f} = \frac{T_{\textrm A} T_{\textrm B}}{T_{\textrm A} + T_{\textrm B}} \ ,
 \label{eqn:Teff}
\end{equation}
where $T_{\textrm A}$ and $T_{\textrm B}$ are the first and second fission transmission coefficients respectively. 
We have considered calculations with a second barrier that is wider and shorter than the first and note that it does not qualitatively change our results. 
Therefore, in the remainder of this work, we only consider the case of a single (maximal) barrier, $T^{(j)}_{\textrm{f}} = T_{\textrm{A}}$. 
Note that for all of the transmission equations above, we have quoted for convinence the `lumped' versions which remove the spin and parity dependence \cite{Mumpower+17}. 

We use the Finite-Range Liquid-Drop Model (FRLDM barrier height predictions from Ref.~\cite{Moller+15} for our primary results. 
Later in section \ref{sec:exp} we study the variation of our predictions based on different barrier heights. 
The FRLDM barrier height predictions have been extensively benchmarked against many nuclear observables including electron-capture delayed fission data, fission-fragment charge yields and a handful of prompt neutron-capture data from weapons tests. 
The predictions of ground state masses relative to these barrier heights shows that there are prominent regions of both (n,f) and \bdf{} for the heavy neutron-rich nuclei near the end of the chart of nuclides with \bdf{} most influential just beyond the $N=184$ closed shell towards the neutron dripline \cite{Moller+15}. 

With the transmission coefficients defined, we can calculate transmission probabilities for these three channels. 
In what follows we simplify the equations by excluding indices of quantum numbers and implicitly take all transitions to obey spin-parity selection rules which changes our transmission coefficients into the so-called `lumped' versions. 
The $\gamma$-emission transition probability in the $j$-th compound nucleus is taken to be
\begin{equation}
 p_j(E_i, E_k) = \frac{1}{N_j(E_i)} T^{(j)}_\gamma(E_i-E_k) \rho_j(E_k)
 \label{eqn:pprob}
\end{equation}
where the transition is from a level with high excitation energy $E_i$ to a level of energy $E_k$, $T^{(j)}_\gamma$ is the $\gamma$-ray transmission coefficient for $c^{(j)}$, $\rho_j(E_k)$ is the level density in $c^{(j)}$ at energy $E_k$ and $N_j(E_i)$ is a normalization factor that we define shortly. 
The neutron-emission transition probability from $c^{(j)}$ to $c^{(j+1)}$ is defined as
\begin{equation}
 q_j(E_i, E_{k^\prime}) = \frac{1}{N_j(E_i)} T^{(j+1)}_n(E_i,E_{k^\prime}) \rho_{j+1}(E_{k^\prime})
 \label{eqn:qprob}
\end{equation}
where the transition is from an energy level $E_i$ in $c^{(j)}$ to an energy level $E_{k^\prime}$ in  $c^{(j+1)}$, $T^{(j+1)}_n$ is the neutron transmission coefficient from $c^{(j+1)}$ to $c^{(j)}$ and $\rho_{j+1}(E_{k^\prime})$ is the level density in $c^{(j+1)}$ evaluated at $E_{k^\prime}$. 
For fission, the transmission probability from an excitation energy $E_i$ below the fission barrier for compound nucleus $c^{(j)}$ is calculated as
\begin{equation}
 r_j(E_i) = \frac{1}{N_j(E_i)} T^{(j)}_f(E_i) \rho^{f}_j(E_i) \ ,
\end{equation}
where $T^{(j)}_f(E_i)$ is the fission transmission coefficient from Eqn.~(\ref{eqn:Tf}) and $\rho^{f}_j(E_i)$ is the fission level density in $c^{(j)}$. 
The normalization factor, $N_j$ is given by the sum over all possible exit channels
\begin{multline}
 N_j(E_i) = \int_0^{E_i} T^{(j)}_{\gamma}(E_i-E_k) \rho_j(E_k) \textrm{d}E_k \\
               + \int_0^{{E_i}-S^{(j)}_n} T^{(j+1)}_n(E_i,E_{k^\prime}) \rho_{j+1}(E_{k^\prime}) \textrm{d}E_{k^\prime} \\
               + \int_0^{E_i} T^{(j)}_{f}(E_k) \rho_j(E_k) \textrm{d}E_k \ ,
 \label{eqn:normprob}
\end{multline}
where the integration in each case runs over the appropriate energy window.
The units of the $q$, $p$ and $r$ quantities are the same as the level density and the transition probabilities do not depend on how the initial state, $E_i$ was populated due to the Bohr independence hypothesis of compound nucleus formation.

The level population in the compound nucleus, $c^{(j+1)}$, at energy, $E_k$, is
\begin{eqnarray}
 \mathscr{P}_{j+1}(E_k) = \sum_{i}^{} \mathscr{P}_{j+1}(E_i) p_{j+1}(E_i, E_k) \nonumber \\
 + \sum_{{k^\prime}}^{} \mathscr{P}_j(E_{k^\prime}) q_j(E_{k^\prime}, E_k) \nonumber \\
 + \sum_{i}^{} \mathscr{P}_{j+1}(E_i) r_{j+1}(E_i) \ ,
 \label{eqn:lvlpop}
\end{eqnarray}
with the summation running over over all levels which may feed the compound state $c^{(j+1)}_k$. 
Each of the terms takes into account the possible pathways of $\gamma$-ray emission, neutron emission and fission to the excited state $E_k$. 
The initial population of this recursive function comes from the $\beta$-decay strength function.

Following from Eqn. (\ref{eqn:lvlpop}), the total production probability (total branching ratio) to emit $j$-neutrons or for the $j$-th compound nucleus to fission are given by,
\begin{subequations}
\begin{eqnarray}
 P_{j\textrm{n}} = \mathscr{P}_{j}(E_k=E_\text{gs}) \text{ where } r_j = 0 \\
 P_{j\textrm{f}} = \sum_{k} \mathscr{P}_{j}(E_k) \text{ where } r_j \neq 0 .\
\end{eqnarray}\label{eqn:Pjs}
\end{subequations}
Thus, for $\beta$-delayed neutron emission, we merely need to calculate the production of the ground state, $E_\text{gs}$ for each compound system, $c^{(j)}$. 
Conversely, for fission, we must sum over all the excited states that end with fission in the particular compound system. 

Equations \ref{eqn:Pjs}a and \ref{eqn:Pjs}b tell us that the statistical decay must end in either the population of one of the daughter generation's ground state or by fission. 
Ergo, the cumulative probability to emit either neutrons or fission must sum to unity
\begin{equation}
 1 = \sum_{j=0}^{10}P_{j\textrm{\small n}} + \sum_{j=0}^{10}P_{j\textrm{\small f}} = P_{\textrm{\small n}} + P_{\textrm{\small f}}
 \label{eqn:cP}
\end{equation}
where the summation index, $j$, represents the number of neutrons emitted and runs over all possible daughter nuclei. 
The value of $P_{j\textrm{n}}$ is the probability to emit $j$-neutrons without fissioning and $P_{j\textrm{\small f}}$ is the probability to fission after $j$-neutrons have been emitted. 
For very neutron-rich heavy nuclei that may participate in the $r$ process, we designate a maximum value of $j=10$. 
The cumulative sum of these two quantities are denoted by $P_{\textrm{\small n}}$ and $P_{\textrm{\small f}}$ respectively. 
Some values of $P_{j\textrm{\small n}}$ or $P_{j\textrm{\small f}}$ may be zero due to selection rules or an exhaustion of the initial excitation energy. 
We explicitly assume that other channels such as proton or alpha emission are very small. 

In the context of the approach outlined above, regular or first-chance \bdf\ is defined as the fission that occurs during the population of the first compound nucleus formation after $\beta$-decay. 
Explicitly, if ($Z$,$A$) is the precursor nucleus that $\beta$-decays, fission occurring in the first generation daughter nucleus, ($Z+1$,$A$), is first-chance \bdf. 
For heavy neutron-rich nuclei there are additional chances to fission after $\beta$-decay stemming from the fact that these nuclei have relatively large $Q_\beta$ values as compared to the fission barriers, $B_\textrm{f}$, or neutron separation energies, $S_\textrm{n}$. 
Thus, neutron-rich nuclei populated in extreme astrophysical conditions may open the possibility for each of the populated daughter generations to fission after $\beta$-decay. 
We use the term multi-chance $\beta$-delayed fission (\mcbdf\ for short) to describe this phenomenon, which is analogous to multi-chance neutron-induced fission that arises at high neutron incident energies. 
In Fig.~\ref{fig:model} this decay mode is represented by fission occuring in the second, or higher, daughter generations. 
If the sum of the fission probabilities after the first daughter generation is greater than 10 \%, $\sum\limits_{j>0}P_{j\textrm{\small f}} > 10 \%$, then we consider this nucleus to undergo \mcbdf{}. 

\begin{figure}
 \begin{center}
  \centerline{\includegraphics[width=95mm]{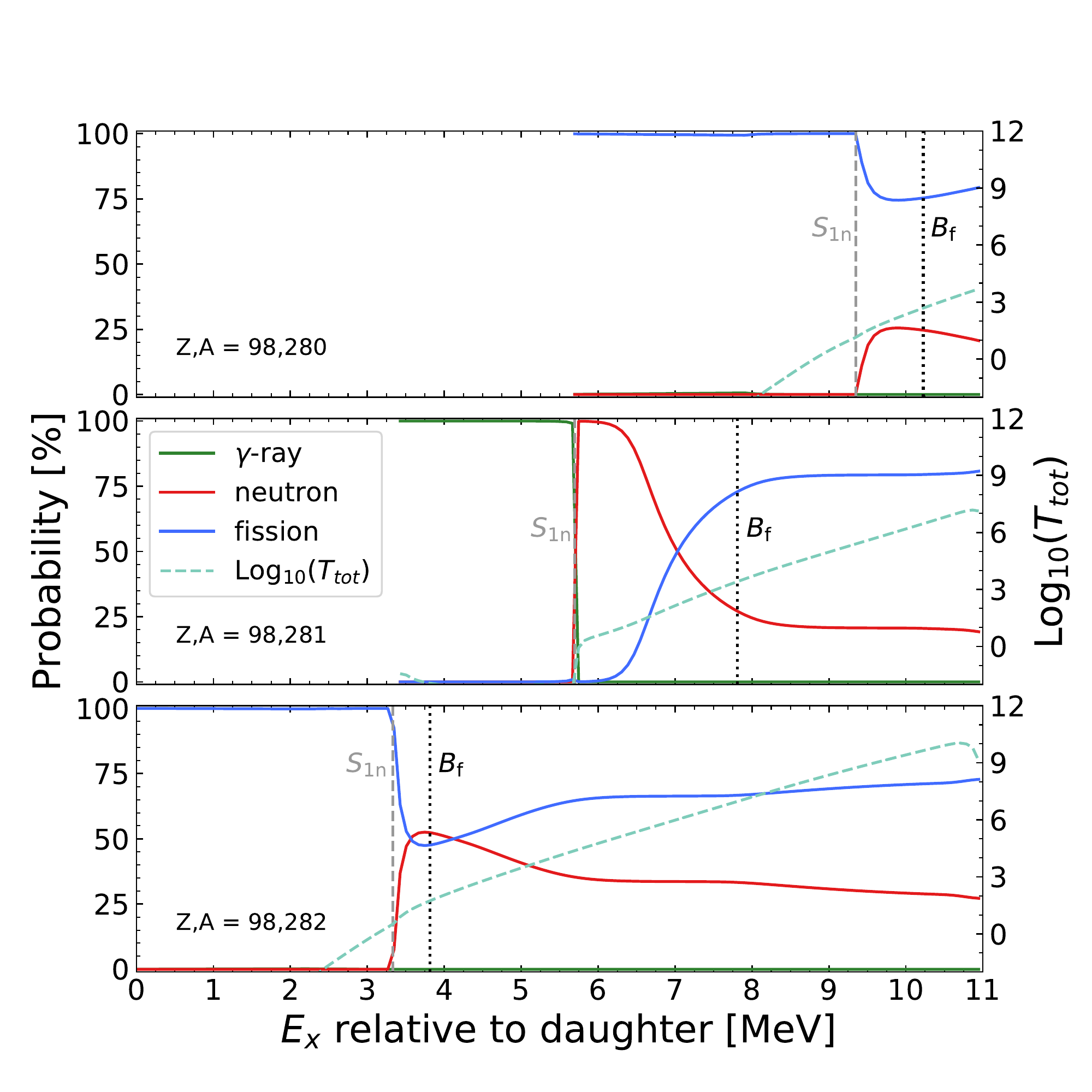}}
  \caption{\label{fig:ngfcomp} (Color Online) Competition between the neutron, gamma and fission channels in the case of beta-delayed fission of precursor nucleus $Z=97$, $A=282$, shown as a ratio of the respective transmission coefficient to the total sum. The total transmission coefficient sum (dashed light line), which spans many orders of magnitude, can be read from the right Y-axis. }
 \end{center}
\end{figure}

Lastly, it is useful in analysis to compute the average neutron multiplicity obtained \textit{after} $\beta$-decay but \textit{pre-scission},
\begin{equation}
 \langle n \rangle = \sum_{j=0}^{10} j \left( P_{j\textrm{\small n}} + P_{j\textrm{\small f}} \right)
\end{equation}
where the summation index again represents the number of neutrons emitted and $P_{j\textrm{\small n}}$, $P_{j\textrm{\small f}}$ are defined in Eqns. (\ref{eqn:Pjs}).

\section{Results}\label{sec:results}
Our results include $\beta$-delayed neutron emission and $\beta$-delayed fission probabilities for all neutron-rich nuclei from stability to extreme neutron excess with an upper limit of the mass number at $A=330$ which represents the extent of the FRDM2012 model. 
We first discuss individual cases before going into the global results of applying the QRPA+HF framework to all nuclei. 

As a typical example, we first explore the \bdf{} of $r$-process nucleus \ce{ ^{282}_{97}Bk }. 
This nucleus has $Q_\beta\sim11$ MeV resulting in a population of high-lying excited states in subsequent daughter generations. 
A competition ensues between the neutron, $\gamma$-ray and fission channels during the statistical decay as shown in Fig.~\ref{fig:ngfcomp}. 
In each successive panel going from bottom to top, the energies are shifted relative to the ground state of the first daughter nucleus. 
For reference, the first daughter generation \ce{ ^{282}_{98}Cf } has $S_\textrm{n}=3.33$ MeV and $B_\textrm{f}=3.82$ MeV while the second and third generations have $S_\textrm{n}=2.36$ MeV, $B_\textrm{f}=4.47$ MeV and $S_\textrm{n}=3.65$ MeV, $B_\textrm{f}=4.54$ MeV respectively. 
The neutron multiplicity is $\langle n \rangle = 1.07$ for this decay chain with large fission chances above $j=0$, thereby satisfying the definition of a nucleus which undergoes \mcbdf.  

Fission dominates the low lying excitations in the first daughter ($j=0$; bottom panel). 
Generally one would expect the $\gamma$-channel to have the largest transmission well below the barrier, however, in some cases such as this one, the channel may be blocked due to selection rules. 
Nevertheless, the total transmission coefficient (right Y-axis) is very small in this energy regime so neither channel contributes to the resultant probabilities. 
Above the neutron separation energy neutron emission is energetically possible, cutting the fission transmission in half since the barrier height is on the order of the separation energy in this nucleus. 
We find that the fission transmission coefficient again increases above the neutron transmission coefficient after about 4 MeV due to the higher fission level density relative to the ground state level density. 

In the second daughter ($j=1$; middle panel), the $\gamma$ transmission coefficient contributes the most to the low excitation energy regime. 
After crossing the neutron separation energy neutrons immediately activates with fission transmission slowly increasing. 
One of the main reasons for the difference between this nucleus and the previous is that the barrier heigh in this nucleus is several MeV higher than the separation energy resulting in a slower onset of fission taking over. 
The third nucleus in the decay chain shows a similar behavior as the first. 

Figure \ref{fig:ngfcomp} hints at a subtle interplay between the $\beta$-strength function, mass surface and fission barrier heights in predicting \bdf\ properties. 
To explore these relationships further we highlight the \bdf\ of $^{295}$Fm and $^{290}$Am. 

\begin{figure}
 \begin{center}
  \centerline{\includegraphics[width=90mm]{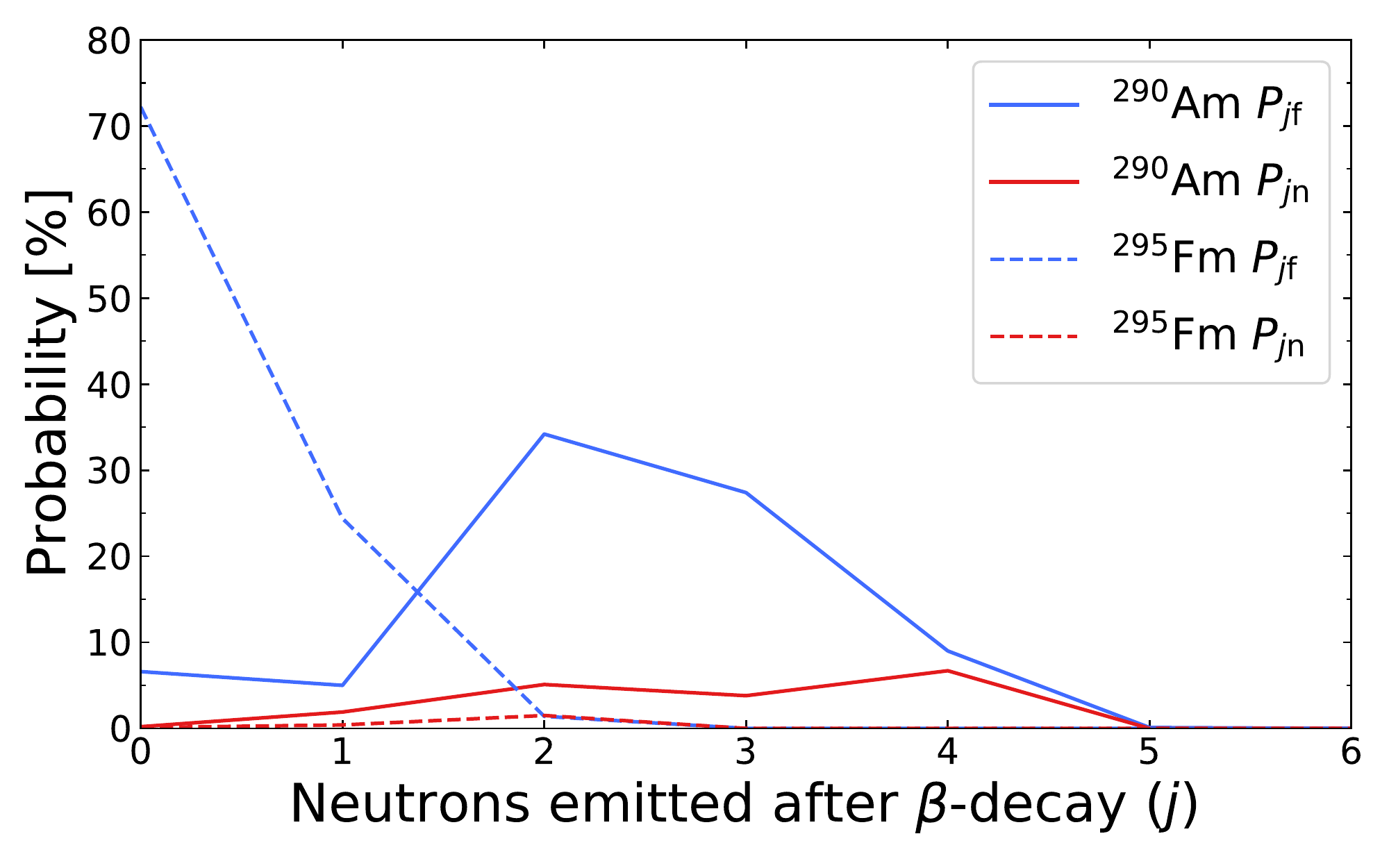}}
  \caption{\label{fig:indprob} (Color Online) Individual probabilities for delayed neutron and delayed fission as a function of $j$ neutron emitted from the daughter nucleus. ${}^{295}$Fm is dominated by fission while ${}^{290}$Am shows more competition between neutron emission and fission. }
 \end{center}
\end{figure}

Figure \ref{fig:indprob} shows the individual $P_{j\textrm{\small n}}$ and $P_{j\textrm{\small f}}$ probabilities after successive neutrons are emitted along the decay chain. 
In the case of \bdf\ of $^{295}$Fm, fission in the first and second daughter generations completely govern the decay. 
The reason for this is that in the first daughter \ce{ ^{295}_{101}Md } has $S_\textrm{n}=3.15$ MeV while $B_\textrm{f}=2.69$ MeV, thus fission operates at low excitation energy and neutron emission is strongly hampered, only reaching a maximum ratio of $25$\% of the total transmission right near the threshold energy. 
The same scenario plays out in the second daughter, thus limiting a longer decay chain with $j>1$. 

Conversely, in the \bdf{} of $^{290}$Am, we see from Fig.~\ref{fig:indprob} that a longer decay chain (up to $j=4$) ensues with significant probabilities for both neutron emission and fission. 
Here the separation energies of the daughter generations are all consistently lower than the respective barrier heights, thus providing ample competition between the neutron and fission channels throughout each isotope participating in the statistical decay. 

\begin{figure}
 \begin{center}
  \centerline{\includegraphics[width=95mm]{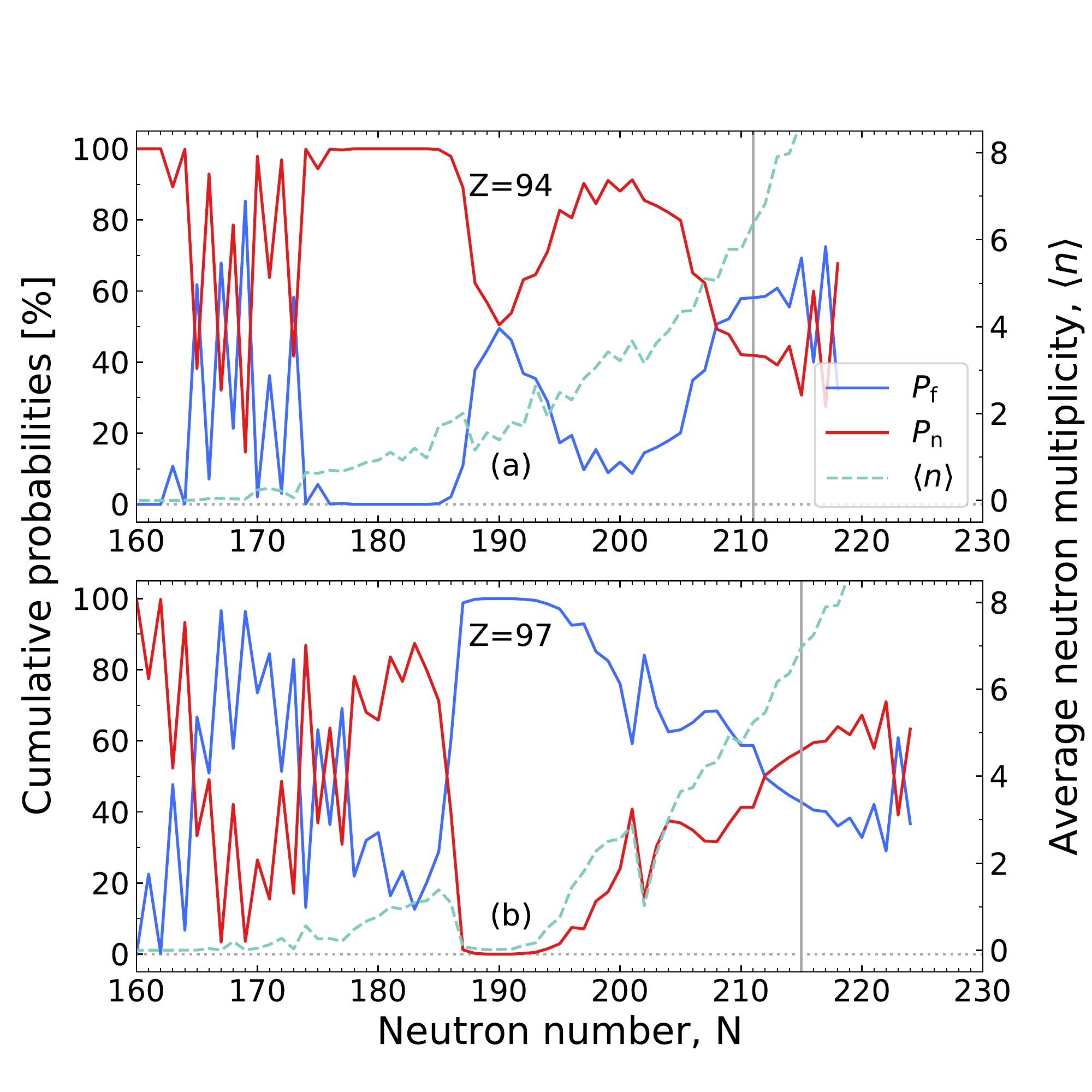}}
  \caption{\label{fig:isocum} (Color Online) (a) The cumulative probabilities for emitting neutrons (blue) or fission (red) after $\beta$-decay for neutron-rich plutonium ($Z=94$) isotopes. The summation of these two terms yields 100\%. The average neutron multiplicity after $\beta$-decay, $\langle n \rangle$, is also displayed by a dashed line and read from the right Y-axis. (b) The same quantities shown for the berkelium ($Z=97$) isotopic chain. Solid grey vertical line indicates the one-neutron dripline in FRDM2012. }
 \end{center}
\end{figure}

Both the cumulative probability to emit a neutron and to fission along an isotopic chain tend to oscillate as shown in Fig.~\ref{fig:isocum} for neutron-rich isotopes of plutonium (a) and berkelium (b). 
The overall trend of the cumulative probabilities is dependent on the fission barrier heights of the nuclei involved in the decay. 
The cumulative \bdf\ probability is small near stability since fission barriers are relatively large here, in agreement with experiments. 
Where fission barriers are relatively low, e.g. near $N=180$ for plutonium and $N=190$ for berkelium, \bdf\ probabilities may reach near 100\%. 
The odd-even staggering seen in the cumulative curves is sensitive to nuclear masses because the difference in nuclear masses sets the threshold energy for neutron emission. 
Starting from stability, the first nucleus in the isotopic chain to have $S_n<0$ is denoted by the solid grey line in both panels. 
This line denotes the extent of the neutron dripline in FRDM2012 that may be accessible in astrophysical applications. 

The average neutron multiplicity before scission, $\langle n \rangle$, is shown on the right Y-axis for the same set of isotopes. 
The increasing value of this quantity with neutron excess shows that \mcbdf\ occupies a substantial amount of real estate after the $N=184$ shell closure. 
Additional neutrons may emitted during scission which we have not considered here that will provide more late-time neutrons for capture in the astrophysical $r$ process \cite{Mumpower+12}. 

\begin{figure*}
 \begin{center}
  \centerline{\includegraphics[width=\textwidth]{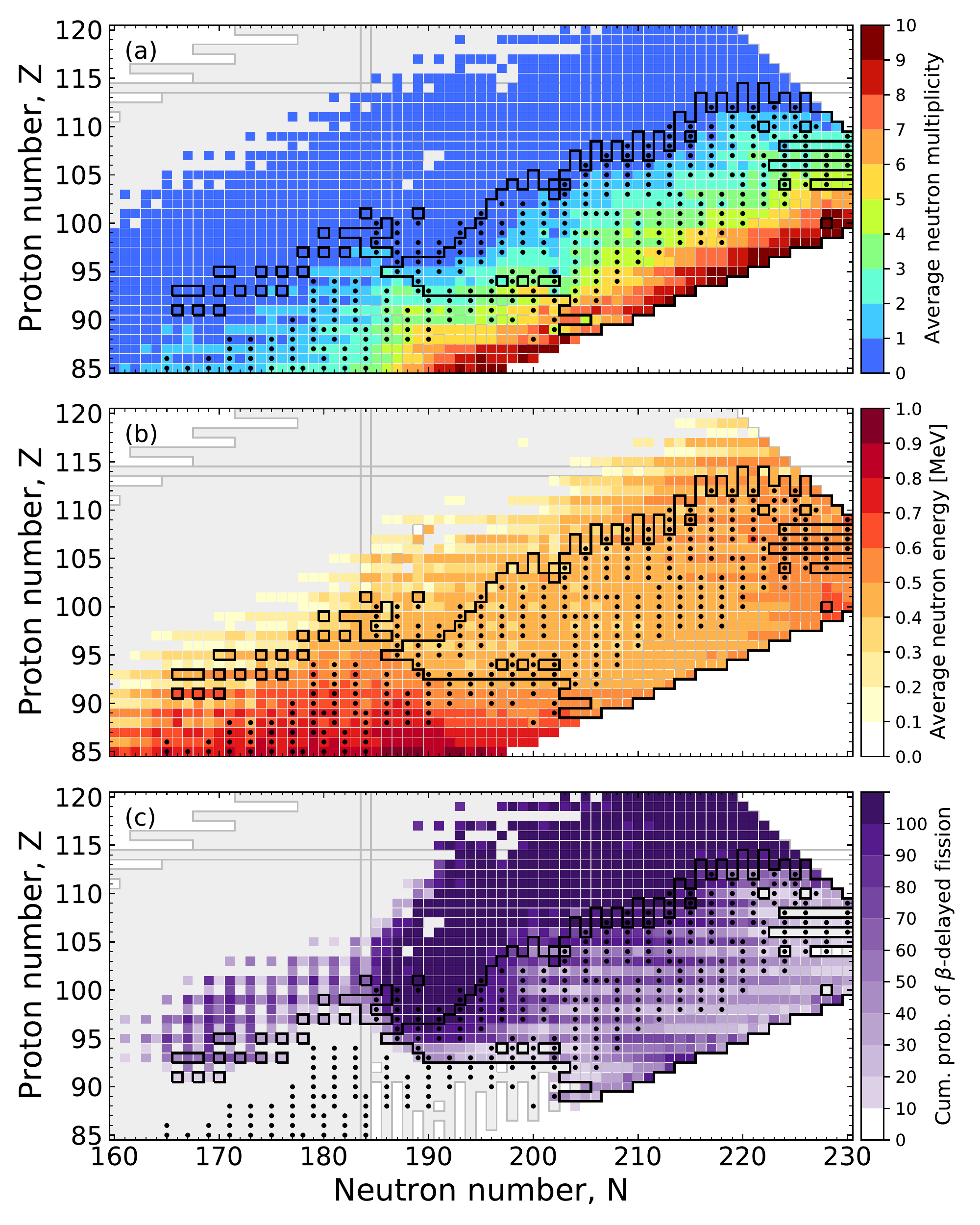}}
  \caption{\label{fig:bdf_results} (Color Online) (a) The average neutron multiplicity, $\langle n \rangle$, after $\beta$-decay for heavy neutron-rich nuclei computed with the QRPA+HF framework. (b) Average $\beta$-delayed neutron emission energy in units of MeV. (c) The cumulative probability for \bdf\ to occur in the region. Nuclei with large cumulative probability and neutron multiplicity are classified as multi-chance \bdf, denoted by solid black border. The black dots denote the $r$-process highway, roughly $S_{\textrm{n}}\sim1$ to $2$ MeV, using FRDM2012 masses. }
 \end{center}
\end{figure*}

We now focus our discussion on the global results shown in Fig.~\ref{fig:bdf_results}. 
The $r$-process highway, nuclei with $S_{\textrm{n}}\sim1$ to $2$ MeV, is denoted by black dots to guide the eye. 
This neutron-rich pathway may extend into the region of large neutron emission as shown in panel (a). 
Towards the neutron dripline roughly five to six neutrons are emitted on average during the decay, similar to other model predictions in the literature \cite{Marketin+16}. 
The average neutron energy after $\beta$-decay is plotted in panel (b). 
Neutron energy for these heavy nuclei is smaller for nuclei that exhibit \mcbdf\ as compared to those surrounding nuclei resulting in a faster thermalization timescale in the context of nucleosynthesis. 

Several additional features of importance for the $r$ process are shown in panel (c) which highlights the cumulative probability for \bdf. 
Beyond the $N=184$ shell closure, many nuclei have $P_{j\textrm{\small f}}=100$\%. 
This means that the $\beta$-decay chain always ends in fission rather than the population of the ground state of any of the daughter generations. 
When this occurs, the nuclear flow of the $r$ process can no longer increase in proton number preventing the production of superheavy elements. 
Nuclear flow may proceed at the dripline, subject to \mcbdf, but network simulations show a termination point around $A\sim300$ \cite{Mumpower+18a}. 
We find this region to extend further than previous calculations \cite{Thielemann+83} suggesting that a neutron-rich pathway to populate superheavies via $\alpha$-decay is unlikely \cite{Meldner+72, Petermann+12}. 
Multi-chance \bdf, outlined by the solid black boundary in Fig.~\ref{fig:bdf_results}, that occurs towards the neutron dripline is another feature of interest for $r$-process nucleosynthesis. 
These nuclei have large $Q_\beta$ and large neutron multiplicity after $\beta$-decay due to small separation energies. 
Unique to this decay process is the observation that the production of light fission products come from a superposition of the yield distributions of the heavy daughter nuclei. 
Details of the impact of first-chance and multi-chance \bdf\ on $r$-process abundances can be found in Ref.~\cite{Mumpower+18a}. 

\section{Nuclei of interest close to stability}\label{sec:exp}
The bulk of the discussion has been centered around nuclei with extreme neutron excess. 
We now turn to the possibility of future experimental measurements probing neutron-rich \bdf{} branching ratios for actinides and superheavies. 
We remind the reader that closer to stability model calculations are more susceptible to deviations, for example from the $\beta$-strength function, than at high neutron excess as has been pointed out in previous work \cite{Moller+03}. 

To isolate potentially interesting nuclei, and reduce dependence on any single model, we account for several variations in our inputs. 
We probe the differences that arise in the $\beta$-strength from three models reported in our previous work \cite{Moller+97, Moller+03, Moller+19}. 
We additionally account for variation in the nuclear mass surface by using various models: the 2012 FRDM \cite{Moller+16}, Duflo-Zuker \cite{Duflo+95}, Hartree-Fock-Bogoliubov (HFB-27) \cite{Goriely+13}, UNEDF1 \cite{Kortelainen+12}, Wood-Saxon \cite{Wang+14} and KTUY05 \cite{Koura+05}. 
Finally, we also consider variation in the estimated fission barrier heights of the models FRLDM \cite{Moller+15}, ETFSI \cite{Mamdouh+98}, HFB-14 \cite{Goriely+09} and KTUY \cite{Koura+14}.
This totals 72 model variations in which we can probe the differences in predicted $\beta$-delayed neutron and \bdf{} probabilities. 
We note that many more additional model variations were performed using other mass models and those results are consistent with and do not alter the conclusions presented below. 

Figure \ref{fig:bdf_exp} shows the resultant model variations for nuclei near stability with the extent of NuBase (2020) in the region shown for reference \cite{NUBASE2020}. 
The colorbar of this figure indicates the percentage of models that predict greater than one percent \bdf{} probability for the given species. 
Near measured nuclei there is little predicted \bdf{} branching. 
However, with increasing neutron excess there is substantial agreement among the models that neutron-rich nuclei undergo at least some \bdf{}. 

From these model variations we form a subset (indicated by the space between NuBase (2020) and the black solid line in Fig.~\ref{fig:bdf_exp}) that may be accessible in future experimental undertakings. 
Our criterion for future accessibility is: (1) neutron-rich nuclei with $90 \leq Z\leq 120$ (2) $P_\mathrm{f}\geq1\%$ and (3) no more than 5 neutrons away from the last neutron-rich value in NuBase (2020). 
This criterion produces a set of 35 nuclei close to stability which are predicted to be most likely to have a measurable $\beta$-delayed fission branching. 
We summarize this information in Table \ref{tab:bdf_exp}. 
The percent models (column 5) in the table represents the number of variations out of 72 total with $P_\mathrm{f}\geq1\%$. 
The minimum fission probability, $P^\textrm{min}_\textrm{f}$ (column 6), represents the smallest $P_\mathrm{f}$ value among the subset of models with $P_\mathrm{f}\geq1\%$ and likewise for the maximum fission probability, $P^\textrm{max}_\textrm{f}$ (column 7).

Several nuclei in this list are only a few neutrons from the last neutron-rich isotope found in the recent evaluated data of NuBase (2020). 
A straightforward way to assess the top experimental candidates is to multiply the percentage of models in agreement of \bdf{} with the minimum \bdf{} probability, $P^\textrm{min}_\textrm{f}$. 
This information can be found in columns 5 and 6 of Table~\ref{tab:bdf_exp}. 
From this estimate one finds primarily superheavies with $^{282}$Bh ($Z=107$) topping the list, but such nuclei may be rather hard to measure owing to low production.
In the actinide region, the odd-$Z$ isotopic chains are of interest with Am ($Z=95$), Bk ($Z=97$), Es ($Z=99$), Md ($Z=101$) and Lr ($Z=103$) hosting several candidates respectively. 
These nuclei may be reachable in the future with fusion evaporation, transfer reactions, or similar techniques. 

\begingroup
\begin{table}
\caption{\label{tab:bdf_exp} Nuclei with \bdf{} branching ratios greater than 1\% ($P_\mathrm{f}=P_{0\mathrm{f}}\geq1\%$) among 72 model variations (see text for details). These nuclei represent prime experimental candidates and they reside no more than 5 neutrons away the most neutron-rich isotope in NuBase (2020). We limit the element number by $80\leq Z\leq 120$.}
\setlength{\tabcolsep}{6pt} 
\renewcommand{\arraystretch}{1.15} 
\begin{tabular}{ c c c c c c c }
\hline
Symbol & $Z$ & $N$ & $A$ & $\%$ Models & $P^\textrm{min}_\textrm{f}$ & $P^\textrm{max}_\textrm{f}$ \\ 
\hline
Pa &  91 & 153 & 244 &  14 &   1 &   4 \\
Pa &  91 & 155 & 246 &  25 &  11 &  26 \\
Np &  93 & 155 & 248 &  14 &   2 &   3 \\
Np &  93 & 157 & 250 &  35 &   1 &  36 \\
Am &  95 & 157 & 252 &   8 &   1 &   3 \\
Am &  95 & 159 & 254 &  75 &   4 &  25 \\
Bk &  97 & 159 & 256 &  58 &   2 &  32 \\
Bk &  97 & 161 & 258 &  75 &   7 &  72 \\
Bk &  97 & 162 & 259 &  18 &   1 &   5 \\
Cf &  98 & 163 & 261 &  17 &   1 &   2 \\
Es &  99 & 163 & 262 &  56 &   1 &  17 \\
Es &  99 & 164 & 263 &   4 &   1 &   1 \\
Md & 101 & 163 & 264 &   8 &   1 &   2 \\
Md & 101 & 165 & 266 &  50 &  41 &  84 \\
Md & 101 & 166 & 267 &  47 &   1 &  28 \\
No & 102 & 167 & 269 &  39 &   1 &  12 \\
Lr & 103 & 165 & 268 &  50 &   4 &  85 \\
Lr & 103 & 166 & 269 &  21 &   2 &   6 \\
Lr & 103 & 167 & 270 &  50 &  90 & 100 \\
Lr & 103 & 168 & 271 &  50 &   2 &  99 \\
Rf & 104 & 167 & 271 &  17 &   2 &  63 \\
Rf & 104 & 169 & 273 &  17 &  59 & 100 \\
Db & 105 & 167 & 272 &  17 &  91 &  94 \\
Db & 105 & 168 & 273 &  14 &  95 & 100 \\
Db & 105 & 169 & 274 &  17 &  94 &  95 \\
Db & 105 & 170 & 275 &  17 &  99 & 100 \\
Bh & 107 & 173 & 280 &  25 &  86 &  94 \\
Bh & 107 & 174 & 281 &  14 &  48 & 100 \\
Bh & 107 & 175 & 282 &  75 &  87 & 100 \\
Bh & 107 & 176 & 283 &  62 &   1 & 100 \\
Hs & 108 & 177 & 285 &  12 &  38 & 100 \\
Mt & 109 & 175 & 284 &  25 &  74 &  93 \\
Mt & 109 & 176 & 285 &   4 &  96 &  99 \\
Mt & 109 & 177 & 286 &  17 &  86 &  94 \\
Mt & 109 & 178 & 287 &  25 & 100 & 100
\end{tabular}
\end{table}
\endgroup

\begin{figure}
 \begin{center}
  \centerline{\includegraphics[width=95mm]{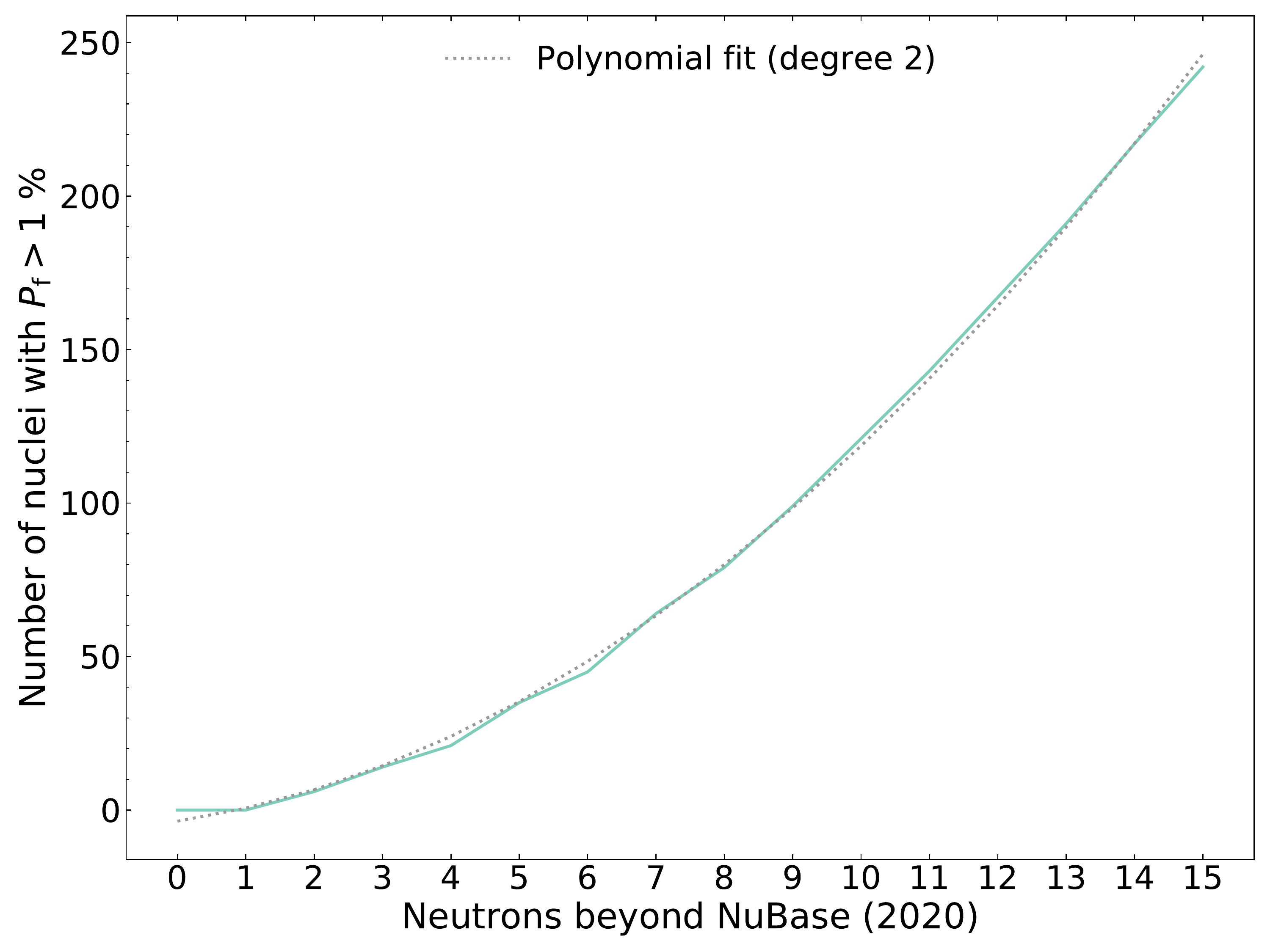}}
  \caption{\label{fig:bdf_counting} (Color Online) Using model variations, the number of nuclei with at least 1 \% chance for \bdf{} increases quadratically (dotted gray) with neutrons beyond NuBase (2020) (green). See text for details. }
 \end{center}
\end{figure}

Further reinforcement that future experimental campaigns may reach measurable \bdf{} branchings comes from the observation that the number of nuclei with $P_\textrm{f}>1$\% increases quadratically with increasing neutron number beyond what has been measured. 
This behavior arises from the above model variations and is shown in Fig.~\ref{fig:bdf_counting}. 
The functional form can be written as
\begin{equation}
 \label{eqn:bdf_counting}
 C_{P_\textrm{f}>1\%} \approx 0.8866 \times n^2 + 3.355 \times n - 3.618 \ ,
\end{equation}
where the coefficients arise from a fit to the observed trend and $n$ measures the distance from the maximum neutron number in NuBase (2020). 
The neutrons counted beyond NuBase (2020) should be $n \geq 1$ when using this equation, owing to the negative value of the last coefficient. 
While $n$ implicitly depends on proton number, the ensemble averaged quantity, represented by $C_{P_\textrm{f}>1\%}$, is insensitive to it. 
This finding comes from the increase of $Q_{\beta}$ values, the decrease of $S_{1\mathrm{n}}$, and rather smooth behavior of predicted fission barriers along isotopic chains. 
With sufficient enough neutron excess, this observed behavior will break down as the predicted $N=184$ shell closure is breached. 
Nevertheless, Equation \ref{eqn:bdf_counting} provides a simple picture for how close experimental efforts are to probing significant delayed fission probabilities. 

\begin{figure}
 \begin{center}
  \centerline{\includegraphics[width=95mm]{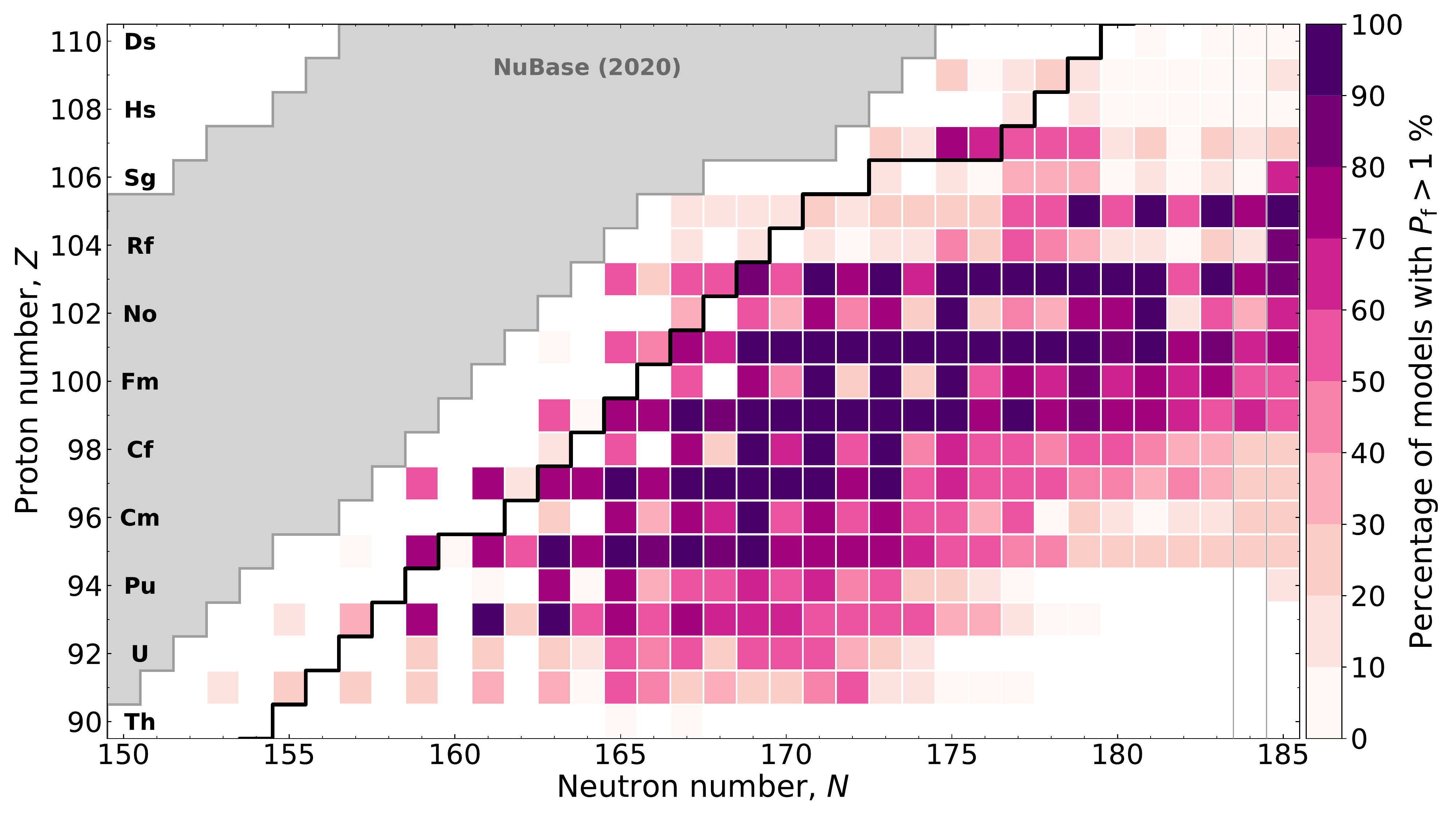}}
  \caption{\label{fig:bdf_exp} (Color Online) The percentage of models that predict \bdf{} with $P_\textrm{f}>1$\% just outside the range of current measurements (see text for details). The grey region indcates the extent of the NuBase (2020) evaluation and the black line is 5 neutrons from it. Shaded nuclei in region between these two bounds represent the first potential candidates for future exploration; these nuclei are summarized in Table \ref{tab:bdf_exp}. }
 \end{center}
\end{figure}

\section{Summary}
We have provided the theoretical underpinnings of the Los Alamos Quasi-particle Random Phase Approximation plus Hauser-Feshbach (QRPA+HF) approach to $\beta$-delayed fission (\bdf{}). 
This approach seeks to combine microscopic nuclear structure information with statistical Hauser-Feshbach theory allowing for a description of the competition between neutrons, $\gamma$-rays and fission during nuclear $\beta$-decay. 

We have calculated $\beta$-delayed neutron emission and $\beta$-delayed fission probabilities for all neutron-rich nuclei from near stability to extreme neutron excess using this framework. 
We find a large region of nuclei which exhibit a propensity for \bdf{}. 
A subset of these extremely neutron-rich nuclei have the potential for multi-chance \bdf{} where a cascade of neutron emission and subsequent fission may occur in each daughter generation after the parent nucleus $\beta$-decays. 

We use variations in our model inputs ($\beta$-strength distribution, mass model and fission barriers) to study the likelihood that future experimental campaigns may observe a measurable \bdf{} branching. 
We show that the number of nuclei with \bdf{} branching increases quadratically with neutron excess relative to current experimental reach. 
From this observed behavior we are hopeful that future endeavors will reveal exciting insights into the nature of delayed neutron and fission branchings among heavy and superheavy elements. 

\acknowledgments
The authors thank Ani Aprahamian for encouraging the initial development of this work. 
The authors would also like to thank Arnie Sierk, J{\o}rgen Randrup, Sean Liddick and Nick Esker for useful discussions. 
The authors finally thank P.~M{\"o}ller for his useful comments and the QRPA $\beta$-strength used in this work. 
This work was carried out under the auspices of the National Nuclear Security Administration of the U.S.  Department of Energy at Los Alamos National Laboratory. 
LANL is operated by Triad National Security, LLC, for the National Nuclear Security Administration of U.S.\ Department of Energy (Contract No.\ 89233218CNA000001).
M.M, T.K. and T.S. were supported in part by the U.S. Department of Energy under Contract No. DE-AC52-07NA27344 for the topical collaboration Fission In R-process Elements (FIRE).

\appendix
\section{Supplemental Data}
We provide the calculated $\beta$-delayed neutron emission and $\beta$-delayed fission probabilities in ASCII format which can be used in suitable applications. 
We additionally provide a copy of Table \ref{tab:bdf_exp} in ASCII format for ease of parsing. 
All of our calculated results can be found online in association with this submission and is released by Los Alamos with number: LA-UR-21-31670.  
Additional model variations can be obtained from the authors via email. 

\bibliographystyle{unsrt}
\bibliography{refs}

\end{document}